# Measurement of coupling development level of new infrastructure investment and digital transformation and its temporal and spatial evolution trend

Sanglin Zhao[1*]; Jikang Cao[2]

1: School of Engineering Management, Hunan University of Finance and Economics, Changsha, China, 410205, 1654952984@qq.com
2: School of Engineering Management, Hunan University of Finance and Economics, Changsha, China, 410205, 2804954288@qq.com
\*Corresponding author：Sanglin Zhao; 1654952984@qq.com

## Abstract

Based on the coupling mechanism between new infrastructure investment level and digital transformation level, a comprehensive index system is constructed. By using entropy weight method, coupling coordination evaluation model, exploratory spatial data analysis (ESDA) and standard deviation ellipse model, the coupling coordination development level of new infrastructure investment level and digital transformation level in 31 provinces and cities in China from 2014 to 2021 is calculated, and its spatial agglomeration level and spatial-temporal evolution characteristics are analyzed. The results show that: ① the comprehensive index of new infrastructure investment level and digital transformation level shows an increasing trend, and the overall digital transformation level shows a gradient decreasing distribution pattern from east to west; ② The coupling coordination between the new infrastructure investment level and the digital transformation level in various provinces and cities is increasing, and more and more provinces are at the basic coordination level or above, but more than half of the provinces are out of balance. Finally, combined with the location distribution and evolution process of the coupling coordination degree between the new infrastructure investment level and the digital transformation level in four major regions of China, the corresponding policy suggestions are put forward.

# 1. Introduction

New infrastructure is a means to actively respond to the new development concept put forward by the state, with the development of network information as the cornerstone and the innovation of science and technology as the driving force, to meet the needs of high-quality development, and to provide infrastructure for the combination of innovation, intelligent upgrading and digital transformation. New infrastructure investment is a new way to invest in new infrastructure, such as 5G, Internet platform, artificial intelligence, etc. It aims to promote the development of the country, expand the needs of society, improve people's happiness in driving emerging industries, ensure social security, and play an important role in promoting rapid and high-quality economic development. The "14th Five-Year Plan" for new infrastructure in 2021 and the National Development and Reform Commission in 2023 further deepened the importance and enthusiasm of new infrastructure investment. Digital transformation aims at creating a high-quality transformation of a new business model, and through further contact with the core business in the company's annual report under the background of some basic digitalization, the goal can be achieved. Because of this, we should intensify the construction of new infrastructure, pay more attention to the investment in new infrastructure, increase the investment in all fields, and expand the development of new infrastructure in all fields, which will build a powerful platform for the development of digital transformation in all fields, lay a great foundation for promoting the leap-forward development of digital transformation and provide a strong means for the rapid development of China's economy.

At present, under the background of economic development entering a new stage, digital transformation is the main trend to promote social and economic development. Investing in new infrastructure and digital transformation are important conclusions drawn by our party in combination with the current economic development stage and development characteristics. In 2022, the "Government Work Report" proposed that in order to promote the digital transformation of industry, the state proposed to build the infrastructure of digital information, which gradually promoted the rapid development of the industrialized Internet, promoted the extensive application of 5G scale, and built an integrated digital center system, thus better empowering economic development and enriching people's lives. It can be seen that under the condition that the mechanism and influencing factors between new infrastructure investment and digital transformation are still unclear, it is of practical significance to study the coupling relationship between new infrastructure investment and digital transformation and its influencing factors and put forward targeted policy suggestions. This paper measures and analyzes the coupling and coordination relationship between new infrastructure investment and digital transformation by provinces and regions by using the relevant index data from 2014 to 2021, and puts forward targeted optimization paths. The first part of the article introduces the background and significance, relevant literature and the innovation of the article. The second part constructs the theoretical framework of the coupling between new infrastructure investment and digital transformation. The third and fourth parts are empirical analysis, in which the coupling coordination degree between new infrastructure investment and digital transformation index system is calculated and analyzed. The fifth part is the last part of the article, which mainly summarizes the empirical results obtained in the article and puts forward the optimization path.

1. Digital transformation

Digitalization has penetrated into every industry in social life, and digital transformation is an urgent and important step to promote social progress. Many domestic scholars have made profound research on digital transformation. Zhang Xun et al. (2019)[1]The better the development of digital inclusive finance, the faster the economic growth, and vice versa. Yang Deming et al (2018) and Shen Guobing et al (2020)[2][3]The influence of digital transformation on the performance of enterprises is studied, but a complete and comprehensive micro-data model is not constructed to discuss digital transformation. Lin Huiyan et al. (2021)[5]It is believed that the vocabulary in the annual report of enterprises can measure the digitalization of listed companies, from which the concept of business conditions of enterprises can show the development path of enterprises. Yao Weighted et al. (2020)[4]The annual reports of listed companies are studied, which fully shows the business information and operating conditions of the company and the management of the enterprise by the management, which has a better reference for the development of the enterprise. Therefore, the digital transformation data in this paper uses company-level data, that is, the average of the provinces where the company is located is converted into provincial data.

2. New infrastructure investment

The investment in new infrastructure is the hub connecting life and work, the important cornerstone of promoting social progress, and the promotion of

A stepping stone to the national governance system and governance capacity. On this basis, many scholars have studied the investment in new infrastructure, and put forward effective suggestions for the development of the country based on the results of this research. Jiang Wei et al. (2023)[6]The relationship between fiscal policy, new infrastructure and high-quality economic development is studied, that is, the higher the investment in new infrastructure, the higher the social investment and consumption of consumers. Zhou Jian (2022)[7]This paper studies the development path of urban digital transformation from the perspective of new infrastructure. Zhou Jia et al. (2022)[8]This paper studies the digital transformation of industrial enterprises driven by new infrastructure, that is, the new and old kinetic energy is driven by new infrastructure, which has promoted the digitalization of industrial enterprises.

### 3. Coupling effect

At present, there are many literatures on the coupling effect between new infrastructure and related indicators, including Dong Yanmei's (2022) research on the measurement and temporal and spatial evolution of the coupling and coordinated development level of new infrastructure investment and scientific and technological innovation ability; Wu Xianfu et al. (2021)[15]Study on the measurement of the coordinated development of new infrastructure construction and strategic emerging industries and its coupling mechanism; Xu Weixiang et al. (2022)[16]Research on the coupling and coordination of new and traditional infrastructure construction in terms of time and space pattern, regional differences and driving factors, etc., but at present, there is no authoritative literature on the coupling relationship between new infrastructure investment and digital transformation. Therefore, based on the important relationship between new infrastructure investment and digital transformation, it is of great significance to study their coupling relationship and form a useful supplement to the existing literature.

## 2. Research methods

### 2.1 Evaluation indicators

The construction of new infrastructure mainly includes three aspects, namely, information infrastructure construction, innovation infrastructure construction and integration infrastructure construction. Among them, the construction of information infrastructure is based on the new generation of information technology, refer to Fan Hejun (2022).[10]Zhao Xing (2022)[11]In this paper, five proxy variables are selected, including domain name number, mobile phone base station, long-distance optical cable line length, mobile phone penetration rate and Internet broadband access port. Innovative infrastructure construction mainly refers to the innovative construction of industrial science and technology integrating production with education, which is referred to Yingjie (2021).[12]In this study, three proxy variables are selected: the number of national university science parks, the number of technology business incubators and the number of torch characteristic industrial bases. The construction of converged infrastructure mainly refers to the popularization of modern transportation and informatization, refer to Gu Bin (2022) and Sanglin Zhao(2024).[9][17]In this study, five proxy variables are selected, including railway mileage, highway mileage, number of computers used by every hundred people, number of websites owned by every hundred enterprises and e-commerce sales. The time range of the above variables is 2014-2021.

Table 1 Evaluation System of New Infrastructure Investment Level

| general objective for | Primary index | Secondary index | unit | Indicator direction |
|---|---|---|---|---|
| Investment level of new infrastructure | Information infrastructure construction | Number of domain names | Ten thousand | + |
| | | Mobile telephone base station | Ten thousand | + |
| | | Long-distance optical cable line length | kilometre (km) | + |
| | | Mobile phone penetration rate | Department/100 people | + |
| | | Internet broadband access port | Ten thousand | + |
| | Innovative infrastructure construction | Number of national university science parks | individual | + |
| | | Number of incubators of science and technology enterprises | individual | + |
| | | Number of characteristic industrial bases of torch | individual | + |
| | Integrated infrastructure construction | operational mileage of railways | kilometre (km) | + |
| | | Graded highway mileage | kilometre (km) | + |
| | | Number of computers used per 100 people | platform | + |
| | | Number of websites per 100 enterprises | individual | + |
| | | Electronic commerce sales | hundred million yuan | + |

In the aspect of digital transformation, from the perspective of enterprises, through the annual report of A-share listed companies from 2014 to 2021, text segmentation is carried out through jieba library, and the word frequencies related to digital transformation are counted, and the mapping relationship between companies and provinces is used to finally summarize them to the provincial level to form the provincial digital transformation index.

Digital Transformation Word Frequency Reference Zhao Chenyu (2021)[13]Based on the processing method, 99 digital related word frequencies in four dimensions: digital technology application, internet business model, intelligent manufacturing and modern information system are counted.

### 2.2 Model building

#### 2.2.1 Entropy Method

Entropy method is an objective weighting method, which can objectively and truly reflect the information implied in the index data. The calculation formula is as follows:

(1) standardized processing:

$$Z_{\theta,ij} = \frac{X_{max} - X_{011}}{X_{max} - X_{min}}$$

(1)

$$Z_{\theta,ij} = \frac{X_{011} - X_{min}}{X_{max} - X_{min}}$$

(2)

In the formula (1) and (2), $Z_{\theta ij}$ is the standardized value of J index in the I province of θ year, $X_{min}$ is the minimum value of index, and $X_{max}$ is the maximum value of index.

(2) calculate the proportion of the j index in the θ year:

$$S_{\theta,ij} = \frac{Z_{\theta,ij}}{\sum_{\theta=1}^{V}\sum_{i=1}^{m} Z_{\theta,ij}}$$

(3)

③ Calculate the information entropy of the j-th index:

$$\mathbf{E}_1 = \left(-\frac{1}{\ln \mathbf{vm}}\right)\sum_{\theta=1}^{v}\sum_{i=1}^{m}\left(\mathbf{S}_{\theta,ij} \times \ln \mathbf{S}_{\theta,ij}\right)$$

(4)

④ Calculate the weight of the j-th index:

$$D_j = 1 - E_j$$

(5)

$$W_j = \frac{D_j}{\sum_{j=1}^{n} D_j}$$

(6)

⑤ Calculate the new infrastructure investment level of the I province in the θ year of each subsystem:

$$f(x)org(y) = \sum_{j=1}^{n} W_j \times \mathbb{Z}_{\theta,j}$$

(7)

In Formula (3)- Formula (7), y is the number of years, m is the number of provinces, n is the number of indicators, $f(x)$ is the investment level of new infrastructure, $g(y)$ is the level of digital transformation.

### 2.2.2 Coupling coordination model

The concept of coupling is used to characterize the cooperative and interactive relationship between two or more systems. This paper analyzes the relationship between new infrastructure investment and digital transformation by using the coupling coordination model, which is expressed as:

$$D = C \times T \tag{8}$$

Among them:

$$C = 2\left\{\frac{f(x) \times g(y)}{[f(x) + g(y)]^2}\right\}^{\frac{1}{2}} \tag{9}$$

$$T = \alpha f(x) + \beta g(x) \tag{10}$$

D is the coupling coordination degree, C is the coupling degree, $f(x)$ represents the new infrastructure investment level after 0-1 standardization, $g(y)$ represents the digital transformation index after 0-1 standardization, T is the comprehensive development index of two subsystems, and α and β are undetermined coefficients. After comprehensive consideration, this paper selects α=β=0.5. On the basis of referring to the existing research, the coupling coordination degree interval of the two systems is divided in the form of Table 2.

Table 2 Division of Coupling Coordination Degree

| Coupling coordination degree | Coordination stage |
|---|---|
| 0≤D≤0.2 | Serious maladjustment |
| 0.2<D≤0.4 | Moderate disorder |
| 0.4<D≤0.5 | Basic coordination |
| 0.5<D≤0.8 | Moderate coordination |
| 0.8<D≤1 | Highly coordinated |

### 2.2.3. Exploratory Spatial Data Analysis (ESDA)

Exploratory spatial data analysis (ESDA) is a collection of spatial data analysis methods and techniques, including the construction of spatial weight matrix, the measurement of global spatial autocorrelation and local spatial autocorrelation, and the identification of spatial correlation.

In this study, the global Moran'sI index and LISA cluster diagram are used to explore whether there is spatial autocorrelation in regional data and identify spatial agglomeration types. The calculation formula is as follows:

$$\text{Moran's I} = \frac{n \sum_{i=1}^{n} \sum_{j=1}^{m} w_{ij}(x_i - \bar{x})(x_j - \bar{x})}{\sum_{i=1}^{m} \sum_{j=1}^{m} w_{ij} \sum_{i=1}^{m}(x_i - \bar{x})^2} \tag{11}$$

$$\text{LISA} = \frac{n(x_1 - \bar{x}) \sum_{j=1}^{m} w_{ij}(x_j - \bar{x})}{\sum_{j=1}^{m}(x_j - \bar{x})^2} \tag{12}$$

Where n is the number of provinces under study; M is the number of neighboring provinces in I province; Wij is the spatial weight; Xi and xj respectively represent the coupling coordination degree between province I and province J; X- is the average value of coupling coordination.

### 2.2.4. Standard deviation elliptic model

The standard deviation ellipse model is suitable for exploring the spatial distribution characteristics of geographical elements and the evolution process of space-time pattern. The calculation includes the center of gravity of ellipse, the standard deviation of long and short semi-axes, azimuth, etc. The calculation method is as follows:

$$N(X,Y) = \left( \sum_{i=1}^{n} w_i X_i / \sum_{i=1}^{n} w_i , \sum_{i=1}^{n} w_i y_i / \sum_{i=1}^{n} w_i \right) \tag{13}$$

$$\sigma_x = \sqrt{\frac{\sum_{i=1}^{n} \left( w_i \bar{X}_i \cos\theta - w_i \bar{y}_i \sin\theta \right)^2}{\sum_{i=1}^{n} w_i^2}} \tag{14}$$

$$\sigma_y = \sqrt{\frac{\sum_{i=1}^{n} \left( w_i \bar{X}_i \sin\theta - w_i \bar{y}_i \cos\theta \right)^2}{\sum_{i=1}^{n} w_i^2}} \tag{15}$$

$$\theta_{i-j} = n\pi/2 + \arctan\left[ (y_i - y_j)/(x_i - x_j) \right] \tag{16}$$

In the above formula, n is the number of provinces under study; (X,Y) is the weighted average barycenter coordinate; Wi is the weight; σx and σy represent the standard deviation on the X axis and Y axis respectively; θ is the azimuth angle of the standard deviation ellipse, that is, the angle formed by clockwise rotation in the true north direction and the long axis of the standard deviation ellipse.

### 2.3 Data sources

31 provinces, autonomous regions and municipalities in China are selected as research samples, and the time range is from 2014 to 2021. The data of new infrastructure investment level come from Guotai 'an database, statistical yearbooks of provinces and China Statistical Yearbook. The digital text comes from the A-share annual report downloaded by Python code from Juchao. The data of the province where the company is located comes from Wind.

## 3. Measurement and analysis of new infrastructure investment level and digital transformation level

Calculate the level of new infrastructure investment and digital transformation from 2014 to 2021, as shown in Table 3. On the whole country, during the period of 2012-2021, China's new infrastructure investment level and digital transformation level both kept rising, from 0.271 and 0.031 in 2014 to 0.394 and 0.158 in 2021 respectively, which reflected the steady development trend of China's new infrastructure investment level and digital transformation level.

Table 3 Comprehensive evaluation results of China's new infrastructure investment level and digital transformation level

| age | 2014 | 2015 | 2016 | 2017 | 2018 | 2019 | 2020 | 2021 |
|---|---|---|---|---|---|---|---|---|
| Investment level of new infrastructure | 0.271 | 0.287 | 0.305 | 0.322 | 0.333 | 0.358 | 0.380 | 0.394 |
| Digital transformation level | 0.031 | 0.046 | 0.061 | 0.078 | 0.089 | 0.104 | 0.128 | 0.158 |

In terms of time series, Sichuan, Guangdong and Shandong rank in the top three in terms of new infrastructure investment, while Guangdong, Beijing and Zhejiang rank in the top three in terms of digital transformation. Combining the two, we can see that Guangdong's new infrastructure investment level and digital transformation level are in the forefront of the country, with obvious advantages.

## 4. Analysis of coupling coordination degree between new infrastructure investment level and digital transformation level

### 4.1 coupling coordination degree time series change analysis

Calculate the coupling coordination between the new infrastructure investment level and the digital transformation level from 2014 to 2021, as shown in Table 4. As can be seen from the table, its mean value, standard deviation and correlation coefficient all show a fluctuating upward trend, rising from 0.252, 0.109 and 0.434 in 2014 to 0.415, 0.282 and 0.439 in 2021 respectively. On the one hand, it reflects that the coupling development of China's new infrastructure investment level and digital transformation level has achieved certain results, on the other hand, it reflects that the absolute difference and relative difference of the coupling coordination degree between the two levels are getting more and more serious.

From the perspective of sub-regions, there are differences in the coupling coordination between the input level of new infrastructure and the level of digital transformation, with the eastern region (0.421) ranking first, followed by the central region (0.375), the northeast region (0.277) and the western region (0.263). From the perspective of different provinces, combined with the calculation results of coupling coordination degree from 2014 to 2021, we can know that: First, Guangdong Province is the only province at a high level of coordination. It is worth mentioning that Guangdong has entered a high level of coordination since 2019, and then its coupling coordination degree has continued to improve steadily. Secondly, in 2021, Jiangsu, Zhejiang, Shandong, Sichuan, Beijing, Hubei, Hunan, Shanghai and Anhui were at a moderate level of coordination, among which Anhui just developed from a basic coordination level to a moderate coordination level in 2021. Based on the above, it can be seen that the coupling coordination degree in the Yangtze River Delta region is also maintained at a high level. These areas are rich in resources, vast in territory, high in economic development level and strong in digital synergy, which is enough to support the extension of the new infrastructure layout. Thirdly, in 2021, Fujian and Hunan are at a basic level of coordination, while Hebei, Jiangxi, Liaoning, Yunnan, Shaanxi, Guizhou, Chongqing, Xinjiang, Guangxi, Heilongjiang, Shanxi, Jilin, Inner Mongolia, Gansu, Tibet and Tianjin are at a moderate level of imbalance. The number of samples in this range is the largest, and the coupling coordination degree of most provinces shows a steady upward trend. Only Shanxi and Inner Mongolia are in a fluctuating period and grow in recent years. Fourthly, in 2021, only Hainan, Qinghai and Ningxia are still in a serious imbalance level, among which Qinghai and Haining are lagging behind due to local digital transformation, while Hainan is mainly limited by geographical location and resource endowment, and the investment level of new infrastructure is low.

On the whole, during the period of 2014-2021, more and more provinces are located at the level of basic coordination and above, especially in 2018, but it should also be noted that more than half of the provinces are out of balance, which shows that the coupling and coordinated development of China's new infrastructure investment level and digital transformation level has indeed achieved results, and there is still much room for improvement.

Table 4 Coupling Coordination Degree of New Infrastructure Investment Level and Digital Transformation Level in Provinces (autonomous regions and municipalities) from 2014 to 2021

| | 2014 | 2015 | 2016 | 2017 | 2018 | 2019 | 2020 | 2021 | average/mean value | sort |
|---|---|---|---|---|---|---|---|---|---|---|
| Shanghai | 0.279 | 0.316 | 0.363 | 0.381 | 0.406 | 0.448 | 0.500 | 0.541 | 0.404 | 9 |
| Yunnan (Province) | 0.241 | 0.248 | 0.252 | 0.270 | 0.284 | 0.305 | 0.339 | 0.367 | 0.288 | 18 |
| Inner Mongolia | 0.197 | 0.213 | 0.242 | 0.261 | 0.258 | 0.269 | 0.267 | 0.288 | 0.250 | 25 |
| Beijing | 0.320 | 0.371 | 0.409 | 0.478 | 0.488 | 0.542 | 0.584 | 0.639 | 0.479 | 6 |
| Jilin (Province) | 0.179 | 0.220 | 0.251 | 0.265 | 0.270 | 0.273 | 0.282 | 0.291 | 0.254 | 24 |
| Sichuan(Province) | 0.368 | 0.419 | 0.447 | 0.488 | 0.512 | 0.547 | 0.592 | 0.626 | 0.500 | 5 |
| Tianjin | 0.061 | 0.123 | 0.130 | 0.131 | 0.148 | 0.154 | 0.181 | 0.209 | 0.142 | 29 |
| Ningxia | 0.086 | 0.093 | 0.105 | 0.114 | 0.117 | 0.113 | 0.122 | 0.132 | 0.110 | 31 |
| Anhui (Province) | 0.295 | 0.325 | 0.369 | 0.393 | 0.412 | 0.433 | 0.469 | 0.512 | 0.401 | 10 |
| Shandong(Province) | 0.399 | 0.443 | 0.476 | 0.500 | 0.525 | 0.538 | 0.594 | 0.626 | 0.513 | 4 |
| Shanxi | 0.204 | 0.266 | 0.274 | 0.293 | 0.306 | 0.304 | 0.299 | 0.302 | 0.281 | 19 |
| Guangdong | 0.564 | 0.627 | 0.686 | 0.749 | 0.798 | 0.834 | 0.876 | 0.933 | 0.758 | 1 |
| Guangxi | 0.199 | 0.231 | 0.262 | 0.272 | 0.288 | 0.300 | 0.306 | 0.327 | 0.273 | 21 |
| Xinjiang | 0.202 | 0.219 | 0.268 | 0.280 | 0.283 | 0.308 | 0.328 | 0.342 | 0.279 | 20 |
| Jiangsu(Province) | 0.400 | 0.460 | 0.499 | 0.540 | 0.565 | 0.601 | 0.646 | 0.699 | 0.551 | 2 |
| Jiangxi | 0.235 | 0.269 | 0.286 | 0.303 | 0.308 | 0.347 | 0.376 | 0.389 | 0.314 | 15 |
| Hebei | 0.233 | 0.265 | 0.286 | 0.313 | 0.334 | 0.352 | 0.371 | 0.390 | 0.318 | 13 |
| Henan(Province) | 0.294 | 0.329 | 0.366 | 0.375 | 0.402 | 0.413 | 0.432 | 0.462 | 0.384 | 12 |
| Zhejiang(Province) | 0.400 | 0.446 | 0.485 | 0.527 | 0.551 | 0.593 | 0.643 | 0.683 | 0.541 | 3 |
| Hainan | 0.068 | 0.090 | 0.098 | 0.115 | 0.122 | 0.132 | 0.138 | 0.169 | 0.117 | 30 |
| Hubei(Province) | 0.360 | 0.405 | 0.425 | 0.436 | 0.452 | 0.477 | 0.504 | 0.538 | 0.450 | 7 |
| Hunan | 0.329 | 0.361 | 0.379 | 0.424 | 0.436 | 0.451 | 0.469 | 0.505 | 0.419 | 8 |
| Gansu | 0.175 | 0.203 | 0.217 | 0.229 | 0.241 | 0.254 | 0.264 | 0.275 | 0.232 | 26 |
| Fujian(Province) | 0.294 | 0.323 | 0.351 | 0.393 | 0.409 | 0.422 | 0.449 | 0.471 | 0.389 | 11 |
| Xizang | 0.110 | 0.105 | 0.129 | 0.159 | 0.176 | 0.187 | 0.209 | 0.237 | 0.164 | 28 |
| Guizhou (Province) | 0.252 | 0.272 | 0.296 | 0.311 | 0.317 | 0.329 | 0.338 | 0.353 | 0.308 | 16 |
| Liaoning (Province) | 0.264 | 0.289 | 0.301 | 0.314 | 0.318 | 0.333 | 0.344 | 0.370 | 0.317 | 14 |
| Chongqing | 0.186 | 0.220 | 0.239 | 0.262 | 0.277 | 0.290 | 0.316 | 0.343 | 0.267 | 22 |
| Shanxi(Province) | 0.247 | 0.264 | 0.292 | 0.303 | 0.314 | 0.332 | 0.347 | 0.364 | 0.308 | 17 |
| Qinghai | 0.166 | 0.176 | 0.183 | 0.192 | 0.193 | 0.198 | 0.179 | 0.169 | 0.182 | 27 |
| Heilongjiang Province | 0.209 | 0.237 | 0.248 | 0.259 | 0.258 | 0.265 | 0.290 | 0.307 | 0.259 | 23 |
| the Northeast area | 0.217 | 0.249 | 0.267 | 0.279 | 0.282 | 0.290 | 0.305 | 0.323 | 0.277 | 3 |
| east | 0.302 | 0.346 | 0.378 | 0.413 | 0.435 | 0.462 | 0.498 | 0.536 | 0.421 | 1 |
| midland | 0.286 | 0.326 | 0.350 | 0.371 | 0.386 | 0.404 | 0.425 | 0.451 | 0.375 | 2 |
| western part of the country | 0.202 | 0.222 | 0.244 | 0.262 | 0.272 | 0.286 | 0.301 | 0.318 | 0.263 | 4 |
| average/mean value | 0.252 | 0.285 | 0.310 | 0.333 | 0.347 | 0.366 | 0.389 | 0.415 | | |
| standard deviation | 0.109 | 0.119 | 0.129 | 0.140 | 0.148 | 0.158 | 0.170 | 0.182 | | |

Coefficient of variation: 0.434 0.420 0.415 0.421 0.426 0.428 0.438 084886

## 4.2 Spatial correlation analysis of coupling coordination degree

### 4.2.1 global spatial autocorrelation analysis

Using exploratory spatial data analysis, the global Moran'sI index of coupling coordination between new infrastructure and digital transformation from 2014 to 2021 is calculated, as shown in Table 5. As can be seen from the table, the global Moran'sI index was positive from 2014 to 2021, and the P value in 2020-2021 was less than 0.05, which passed the significance test at 95% confidence level, and the P value in 2014-2021 showed a rising trend. It can be shown that there is a positive spatial correlation between the investment in new infrastructure and the coupling coordination degree of enterprise digital transformation.

Specifically, the overall Moran'sI index of the coupling coordination degree between new infrastructure and digital transformation shows an upward trend. From 2014 to 2016, the overall Moran'sI index of the coupling coordination degree between new infrastructure and digital transformation is on the rise, and the increase rate is the fastest in the calculation year; In 2016-2017, the overall Moran'sI index of the coupling coordination degree between new infrastructure and digital transformation decreased slightly, and in 2018-2021, the overall Moran'sI index of the coupling coordination degree between new infrastructure and digital transformation showed an upward trend again. According to the global Moran'sI index of 8 years, the overall level of coupling and coordination between new infrastructure and digital transformation of enterprises is different from that of provinces.

The clustering characteristics of domains. This is because, as the starting point of steady growth, all provinces and cities have accelerated the construction of new infrastructure, which makes the spatial correlation between new infrastructure and digital transformation show an increasing trend, and also makes the digital transformation of enterprises have stronger support, so the construction of digital information infrastructure will be accelerated.

Table 5 Global Moran'sI Index of Coupling Coordination Degree in 2014-2021

| index | 2014 | 2015 | 2016 | 2017 | 2018 | 2019 | 2020 | 2021 |
|---|---|---|---|---|---|---|---|---|
| Moran'sI | 0.15 | 0.173 | 0.177 | 0.16 | 0.18 | 0.183 | 0.21** | 0.22** |
| P | 0.112 | 0.073 | 0.066 | 0.091 | 0.062 | 0.059 | 0.035 | 0.028 |
| Z | 1.591 | 1.792 | 1.836 | 1.19 | 1.869 | 1.887 | 2.111 | 2.192 |

**4.2.2 Local spatial autocorrelation analysis**

In order to better observe the local spatial difference characteristics of coupling coordination degree, the LISA aggregation map is made by using Geoda and ArcGIS software, as shown in Figure 2. From the LISA agglomeration maps in 2014 and 2021, it can be seen that the coupling coordination degree between new infrastructure and enterprise digital transformation presents a trend of agglomeration, which is embodied in three high-value agglomeration areas in the southeast and west and two low-value agglomeration areas in the northwest. From 2014 to 2021, there were two provinces and cities with changes in agglomeration types, accounting for 6.5% of the sample size. It can be seen that the spatial distribution of coupling coordination degree between new infrastructure and enterprise digital transformation is relatively stable.

As of 2021, Jiangsu, Anhui and Fujian provinces are high-altitude (HH) agglomeration areas, and it can be known that the surrounding provinces and cities are more easily driven by the radiation of several high-altitude provinces and cities along the eastern coast, and the spatial correlation of several coastal provinces shows high-level agglomeration effects; Jiangxi is a low-high-level (LH) cluster, which shows that the level of coupling and coordination between new infrastructure and digital transformation in this region is low, while the development level of neighboring regions is relatively high. Compared with other provinces and cities, Jiangxi's new infrastructure foundation is low, its economic foundation is not high enough, there are relatively few high-quality development enterprises, and the digital transformation of enterprises is difficult, especially in comparison with other provinces and cities with strong economic foundations around it. Xinjiang and Inner Mongolia are low-level (LL) clusters. These two provinces are located in the northwest of China, with poor economic foundation, weak industrial foundation and relatively late and slow infrastructure development, and the new infrastructure has a slightly weak role in promoting the digital transformation of enterprises. Sichuan is a high-low (HL) agglomeration area, and its spatial correlation is characterized by a regional polarization pattern of high in the middle and low in the periphery. Sichuan's overall economy develops rapidly, and Sichuan Basin in the province has natural geographical advantages, which has obvious radiation driving effect on surrounding provinces and cities.

### 4.3 Spatial Pattern Evolution Analysis of Coupling Coordination Degree

**1. Location distribution of spatial aggregation of coupling coordination degree**

Using the standard deviation ellipse model to explore the location distribution of the coupling and coordination relationship between new infrastructure and digital transformation in 2014-2021, the relevant parameters of the standard deviation ellipse in 2014-2021 are measured, as shown in Table 6.

① Change of distribution center of gravity. From 2014 to 2021, the center of gravity of the coupling coordination between new infrastructure and digital transformation is between 112.63 E ~ 112.75 E and 33.03 N ~ 33.17 N, all of which are located in Henan. In 2014-2017, the center of gravity of coupling coordination moved slightly to the northeast at the speed of 3.11km/ year, and in 2017-2021, the center of gravity of coupling coordination moved to the southeast at the speed of 4.06km/ year, and the moving speed increased.

② Change of distribution range. From 2014 to 2021, the standard deviation ellipse area of the coupling coordination degree between new infrastructure and digital transformation showed an overall increasing trend. In 2021, the national standard deviation ellipse area increased by 0.76 million km compared with 2014. Specifically, the national standard deviation ellipse area increased by 28,200 km in 2017 and decreased by 20,500 km in 2021 compared with 2017. In addition, the major axis of the ellipse was reduced from 1113.16km in 2014 to 1110.72km in 2021, and the minor axis increased from 9979.87km in 2014 to 1002.4km in 2021. ③ Changes of distribution shape and direction. Judging from the standard deviation ellipse shape of the coupling coordination degree between new infrastructure and digital transformation in 2014-2021, the overall long semi-axis becomes shorter and the short semi-axis becomes longer. From the direction of the standard deviation ellipse, the standard deviation ellipse of the coupling coordination degree between new infrastructure and digital transformation increased from 51.79 to 54.69 in 2014-2021. It can be seen that the coupling coordination degree of the southwest region of the standard deviation ellipse grew faster than that of the southeast region. According to the above analysis, it can be found that from 2014 to 2021, the coupling and coordination degree between new infrastructure and digital transformation presents an obvious center-periphery structure. From 2014 to 2021, the center of gravity of the ellipse experienced the trend of moving to the northeast first and then to the southeast. On the whole, the longitude moved from 112.63 to 112.75, and the actual movement was 11.18km, and the latitude moved from 33.11 to 33.03, and the actual movement was 8.9km. The moving distance in the longitude direction was slightly larger than that in the latitude direction, indicating that the difference in coupling coordination between east and west was greater than that between north and south. In addition, the area of the ellipse has increased from 348.98 in 2014 to 3,497,600 km in 2021, indicating that the policy measures and actual effects adopted in the coupling development of new infrastructure and digital transformation in China are becoming more and more balanced nationwide. We should continue to innovate new infrastructure development models and increase investment in new infrastructure, and at the same time, make precise efforts in policy promulgation and regional coordinated development according to the actual situation of various provinces and cities to promote the coordinated development of new infrastructure and digital transformation of enterprises in China.

**2. Spatial location distribution of internal coupling coordination degree in four regions**

After the overall location analysis of the coupling coordination degree between new infrastructure and digital transformation in China, in order to further study the temporal and spatial differences and evolution characteristics of the coupling coordination degree between new infrastructure and digital transformation, China is divided into four regions: western region, central region, eastern region and northeast region, and the standard deviation ellipse and the change trend of the center of gravity of the coupling coordination degree between new infrastructure and digital transformation in the four regions are calculated, as shown in Table 6.

Table 6 Standard Deviation Elliptic Parameters of Coupling Coordination Degree between New Infrastructure and Digital Transformation

| zone | age | Longitude of gravity center | Gravity center latitude | Long axis (km) | Short axis (km) | Azimuth (degree) | Area (km) |
|---|---|---|---|---|---|---|---|
| whole country | 2014 | 112.63 | 33.11 | 1113.16 | 997.99 | 51.79 | 348.99 |
|  | 2017 | 112.70 | 33.17 | 1111.24 | 1007.80 | 51.56 | 351.81 |
|  | 2021 | 112.75 | 33.03 | 1110.72 | 1002.40 | 54.69 | 349.76 |
| midland | 2014 | 113.72 | 31.32 | 510.41 | 271.26 | 176.70 | 43.49 |
|  | 2017 | 113.74 | 31.41 | 522.13 | 272.50 | 176.06 | 44.69 |
|  | 2021 | 113.83 | 31.28 | 504.02 | 276.64 | 176.65 | 43.80 |
| western part of the country | 2014 | 102.92 | 32.88 | 1085.39 | 837.54 | 146.21 | 285.57 |
|  | 2017 | 102.85 | 32.99 | 1103.34 | 842.78 | 143.52 | 292.11 |
|  | 2021 | 102.69 | 32.77 | 1103.88 | 840.10 | 141.29 | 291.32 |
| east | 2014 | 117.38 | 31.43 | 975.42 | 356.68 | 6.21 | 109.28 |
|  | 2017 | 117.32 | 31.54 | 994.01 | 361.16 | 6.22 | 112.76 |
|  | 2021 | 117.35 | 31.63 | 997.14 | 366.55 | 6.39 | 114.81 |
| the Northeast area | 2014 | 125.20 | 44.02 | 501.90 | 79.07 | 21.25 | 12.46 |
|  | 2017 | 125.30 | 44.04 | 488.20 | 82.40 | 21.16 | 12.63 |
|  | 2021 | 125.29 | 44.06 | 493.90 | 81.27 | 21.13 | 12.60 |

From 2014 to 2021, the center of gravity of the coupling and coordination between new infrastructure and digital transformation in Northeast China is 125.2.

E ~ 125.3 E, 44.02 N ~ 44.06 N, all located in Jilin Province, moved to the northeast at a speed of 2.77km/ year in 2014-2017, and moved to the northwest at a speed of 0.59km/ year in 2017-2021. The standard deviation ellipse in the northeast showed a northeast-southwest trend, and it was long. The standard deviation ellipse area increased from 124,600 km in 2014 to 126,000 km in 2021, with an increase of 0.14 million km. The standard deviation ellipse azimuth is reduced from 21.25 to 21.13. It can be seen that the standard deviation ellipse in Northeast China, the growth rate of coupling coordination in the northern region is higher than that in the southern region, the coupling coordination tends to increase in the east-west direction, and the inter-provincial gap widens. In the future, it is necessary to further balance the development of various regions, formulate reasonable development policies according to the innate location advantages and acquired economic foundation of each region, and further strengthen the cooperation between government and enterprises within and across regions, develop new technologies, promote the industrialization of industry and agriculture, give play to their agricultural advantages, optimize the design of smart agriculture, and promote the digital transformation of catalytic enterprises by new infrastructure. From 2014 to 2021, the center of gravity of the coupling coordination degree between new infrastructure and digital transformation in the eastern region is 117.32E~117.38°E and 31.43 n ~ 31.63 n, all of which are located in Anhui Province, moving to the northwest at a speed of 4.5km/ year in 2014-2017 and 2.6km/ year in 2017-2021. The standard deviation ellipse in the eastern region presents a northeast-southwest trend, with the long semi-axis increasing from 975.42km in 2014 to 997.14km, and the short semi-axis increasing from 35.67km in 2014 to 36.66km in 2021, showing a flattening trend. The standard deviation ellipse area increased from 1,092,800 km in 2014 to 2021.

1,148,100 km, an increase of 55,300 km. The standard deviation ellipse azimuth increases from 6.21 to 6.39. It can be seen that the coupling coordination degree in the eastern region is gradually expanding in the east-west and north-south directions, and the radiation driving effect of the eastern coastal provinces and cities around the north-south direction is obvious. In the future, it is necessary to continue to promote the leading role of provinces and cities with higher development levels in the east and west, narrow the development gap between the east and west, give full play to regional advantages and technological innovation advantages, further improve the level of urbanization and internationalization by making use of its convenient shipping, continuously amplify the demonstration effect of reform and innovation, and fully release the policy dividend. From 2014 to 2021, the center of gravity of the coupling coordination between new infrastructure and digital transformation in central China is 113.72E~113.83°E, 31.28 n ~ 31.41 n, all of which are located in Hubei, moving to the northeast at a speed of 3.4km/ year in 2014-2017 and 4.2km/ year in 2017-2021. The standard deviation ellipse in the central region presents a northwest-southeast trend, with the major axis reduced from 510.41km in 2014 to 504.02km, and the minor axis increased from 271.26km in 2014 to 276.64km in 2021. The standard deviation ellipse area increased from 435,000 km in 2014 to 438,000 km in 2021, with an increase of 3,000 km. The standard deviation ellipse azimuth increases from 176.7 to 176.65. It can be seen that the coupling coordination degree of new infrastructure and digital transformation in the central region extends horizontally in the east-west direction and slightly contracts in the north-south direction. In the future, the central region needs to play a better role as a bridge, give full play to its own energy advantages, undertake industrial migration in the eastern region, actively carry out new infrastructure transportation construction between the east and the west, strengthen inter-provincial cooperation and exchanges, actively integrate into major regional strategies, deepen cooperative linkage and coordinated development, and pay attention to disasters such as plateau soil erosion, floods and sandstorms brought about by economic development to promote high-quality development. From 2014 to 2021, the center of gravity of the coupling coordination degree between new infrastructure and digital transformation in the western region is 102.69 E ~ 102.92 E and 32.778 N ~ 32.99 N, all of which are located in Sichuan, moving to the northwest at a speed of 4.62km/ year in 2014-2017, and at a speed of 7.7 in 2017-2021. The standard deviation ellipse in the central region presents a northwest-southeast trend, with the major axis increasing from 1085.39km in 2014 to 1103.88km, and the minor axis increasing from 837.54km in 2014 to 840.1km in 2021. The standard deviation ellipse area increased from 2,855,700 km in 2014 to 2,913,200 km in 2021, with an increase of 57,500 km. The standard deviation ellipse azimuth is reduced from 146.21 to 141.29. It can be seen that the spatial agglomeration effect of the coupling and coordination between new infrastructure and digital transformation in the western region is rising. In the future, it is necessary to further strengthen infrastructure planning and construction, optimize the energy supply and demand structure, promote the formation of a modern industrial system, seize the opportunity of the "Belt and Road" development, strengthen the construction of open channels, continuously improve the innovation and development capabilities, and promote the overall balanced development of the western region.

## 5. Conclusions and recommendations

### 5.1 Conclusion

Based on the panel data of 31 provinces and cities from 2014 to 2021, this paper measures the capacity of new infrastructure and digital transformation based on entropy weight method, and analyzes the time series evolution, aggregation characteristics and location distribution of the coupling coordination degree of new infrastructure and digital transformation in 31 provinces and cities by using the coupling coordination degree model, exploratory spatial data analysis (ESDA) and standard deviation ellipse model. The main conclusions are as follows:

On the whole, China's new infrastructure investment level and digital transformation level show an increasing trend, from 0.271 and 0.031 in 2014 to 0.394 and 0.158 in 2021, respectively, indicating that the effectiveness of China's new infrastructure and digital transformation is remarkable. The layout of new infrastructure investment level and digital transformation level generally shows a decreasing trend from east to west, with unbalanced development among regions and great spatial differences.

During the research period, China's provinces at the level of basic coordination and above showed an upward trend, especially in 2018, but more than half of the provinces were out of balance, indicating that the coupled and coordinated development of China's new infrastructure investment level and digital transformation level has indeed achieved results, but there is still much room for improvement.

From 2014 to 2021, there is a positive spatial correlation between the investment in new infrastructure and the digital transformation of enterprises, and the overall Moran'sI index shows an upward trend. According to the global Moran'sI index for eight years, the overall level of coupling and coordination between new infrastructure and digital transformation of enterprises shows regional agglomeration characteristics from the inter-provincial perspective, and the spatial correlation shows an increasing trend.

From 2014 to 2021, the center of gravity of the national standard deviation ellipse of coupling coordination moved to the southeast as a whole, and the standard deviation ellipse was distributed in the northeast and southwest directions. The clustering characteristics of coupling coordination in the north-south direction became more and more obvious, and the gap between the north and the south increased. Specifically, the center of gravity of the standard deviation ellipse in the central region moves to the southeast as a whole, and the quasi-deviation ellipse is distributed in the northwest-southeast, which is characterized by the east-west expansion, the north-south agglomeration and the widening gap between the north and the south. It is necessary to further play a good role as a bridge to strengthen inter-provincial cooperation and exchanges and actively integrate into regional major strategies; The standard deviation ellipse in the western region is northwest-southeast, the center of gravity shifts to the southwest as a whole, and the long and short semi-axes all show an increasing trend, and the aggregation situation and regional differences are reduced. In the future, it is necessary to further strengthen the construction of open large channels and promote high-quality development; The standard deviation ellipse in the eastern region is northeast-southwest, and the center of gravity moves to the northwest as a whole, showing an expansion trend. The high-level development provinces and cities have obvious radiation driving effects on the north and south and the east and west regions, and it is necessary to continue to enlarge the demonstration effect of reform and innovation in the future; The standard deviation ellipse in Northeast China is northeast-southwest, and the center of gravity moves to the northeast as a whole, and the north and south are gathering, indicating that the gap between Heilongjiang, Jilin and Liaoning provinces is expanding.

### 5.2 Suggestions

**1. Give full play to the leading role of government macro-control.**

The government should give guidance by implementing fiscal policy and differentiation. Through the above research, we can know that the spatial difference of new infrastructure layout in China and the degree of digitalization of enterprises in different regions are quite different. Therefore, for the eastern region, we should increase financial input while improving investment efficiency, reduce the waste of funds caused by blind investment, and improve the level of high-quality development. For the areas with low development level in the central and western regions and the northeast region, we should do a good job in overall control layout and development planning, take financial assistance and policy incentives, and cultivate and absorb high-tech enterprises to settle in.

**2. Promote the development of new infrastructure according to local conditions**

From the research, we can see that the new infrastructure in China presents a decreasing spatial layout from the east to the west, and the spatial difference is obvious. The construction, demand and future development potential of new infrastructure facilities in different regions and different provinces and cities are different. The government should combine the industrial and agricultural bases and economic strength of various regions, fully consider the industrial support capacity and regional carrying capacity, and promote the development of new infrastructure in a targeted manner, such as making the developed areas in the east constantly enlarge the demonstration effect of reform and innovation, speeding up the digital transformation of economy and society, giving priority to the development of infrastructure in the underdeveloped areas in the west, and focusing on areas such as people's livelihood on the basis of local development history and economic foundation on the basis of continuous improvement of infrastructure.

### 3. Improve the quality of coupling and coordination between new infrastructure investment and digital transformation.

Organically combine the national and regional perspectives to improve the coupling and coordination between new infrastructure investment and digital transformation. From the national level, the average, standard deviation and coefficient of variation of the coupling coordination between new infrastructure investment and digital transformation in 2014-2021 show that the differences between different regions are becoming increasingly obvious. In the future, it is necessary to further proceed from the whole and strengthen the top-level design based on the overall situation. From the regional perspective, for provinces and cities with high coupling and coordination between new infrastructure investment and digital transformation, on the basis of maintaining the leading edge, we should further refine the policies of new infrastructure investment and promoting digital transformation, promote the cooperation and exchanges between neighboring provinces and provinces, and give play to the radiation-driven role of developed provinces and cities to neighboring provinces and cities. For provinces and cities with low coupling and coordination between new infrastructure investment and digital transformation, we should seize the opportunity of "One Belt, One Road" construction, seize the policy dividend, and actively integrate into major regional strategies and strengthen.

**Disclosure of interests**
All authors disclosed no relevant relationships
**Author contributions**
Sanglin Zhao:Conceptualization, Data curation, formal analysis, Software, Writing–original draft, Visualization, Writing–review and editing;  Jikang Cao:Writing–review and editing
**Data availability**
Data available on request from the Corresponding author.
**Funding statement**

This study did not receive any funding